# Spontaneous Directional Motion of Shaped Nanoparticle

Nan Sheng[1], YuSong Tu[2], Pan Guo[1], RongZheng Wan[1], ZuoWei Wang[3,†], and HaiPing Fang[1,*]

[1] *Division of Interfacial Water and Key Laboratory of Interfacial Physics and Technology, Shanghai Institute of Applied Physics, Chinese Academy of Sciences, P.O. Box 800-204, Shanghai 201800, China*

[2] *College of Physics Science and Technology, Yangzhou University, Jiangsu, 225009, China*

[3] *School of Mathematical and Physical Sciences, University of Reading, Whiteknights, Reading RG6 6AX, United Kingdom*

**In nanoscale space and pico- to nanoseconds enormous physical, chemical and biological processes take place, while the motions of involved particles/molecules under thermal fluctuations are usually analyzed using the conventional theory of diffusive Brownian motion based on both sufficiently long time averaging and assumptions of spherical particle shapes. Here, using molecular dynamics simulations, we show that asymmetrically shaped nanoparticles in dilute solutions possess spontaneous directional motion of the center of mass within a finite time interval. The driving force for this unexpected directional motion lies in the imbalance of the interactions experienced by their constituent atoms during the orientation regulation at timescales before the onset of diffusive Brownian motion. Theoretical formulae have been derived to describe the mean displacement and the variance of this directional motion. Our study potentially takes an important step towards establishing a complete theoretical framework for describing the motions of variously-shaped particles in solutions over all timescales from ballistic to diffusive regime.**

The motion of molecules caused by thermal fluctuations plays an essential role in determining the probability of meeting their targets upon functioning[1–5], such as nucleation of clusters[6], self-assembling[7], triggering chemical reaction[8], intercellular signal transduction[9], neurotransmission[10], and various other physical processes[11,12], chemical reactions[13] and biological functioning[14,15]. Conventionally, the molecules/particles are simply treated as perfect spheres whose trajectories are described as random walks or free diffusion and their mean square displacements (MSD) obey the Einstein relationship.[16–18] With the development of nanoscience and nanotechnology, it has been realized that the motion of particles within spaces comparable to their own sizes or within finite timescales is very distinctive from the conventional theory predictions. Han *et al*. have experimentally observed the anisotropic diffusion behavior of ellipsoidal particles along different axial directions within timescales up to second.[19] Chakrabarty *et al.* have found that the Brownian motion of boomerang colloidal particles of micrometer size confined in two-dimensions (2D) exhibits a biased mean displacement for the cross point of the two arms, which represents the center of body (CoB), towards the center of hydrodynamic stress (CoH).[20,21] On the finite timescale, Huang *et al.* found a slow transition from ballistic to diffusive Brownian motion by experimentally measuring the mean-square displacement of a 1-μm-diameter silica sphere in water using optical trapping technique.[22] We note that, a majority of the chemical[23,24], biological[25,26] and physical[27,28] processes take place in nanoscale space[14,15] and pico- to nanoscale time intervals[29,30], such as chemical reactions and neurotransmission. However, to our knowledge, very few studies have been reported on the motion of molecules under thermal fluctuations both in spaces comparable with their own sizes and within very short time intervals. In a previous work based on molecular dynamics (MD) simulations, we observed spontaneous anisotropic motions of small asymmetric solute molecules, such as methanol and glycine, in water within a finite time, which showed preference in certain specific directions along with the molecular orientations[31,32]. These simulation results imply the existence of spontaneous directional motion of particles solely under thermal fluctuations, which is apparently anti-intuitive and so raises many challenging questions, especially upon the generality and important implications, the underlying physical mechanism and possible theoretical description of such phenomenon. Why does the motion of the centers of mass of the molecules bear a directional behavior and what is the physical origin of the driving force for the directional motion since there is not any external interference? How do the timescales of such directional motion compare with the conventional picture of diffusive Brownian motion?

In this work we tackle these challenges by studying the motion of a model nanoparticle shaped as triangular pyramid in very dilute solution using both three-dimensional (3D) MD simulations and theoretical analyses. Our MD simulations with explicit solvents are carried out by solving dynamic equations based on direct interactions between atoms, and so avoid making any prior assumptions, such as isotropic random fluctuations and isotropic dissipation, as used in the Langevin equation[18]. We show clear evidence that there is a spontaneous directional motion process of the center of mass (CoM) of the shaped nanoparticle and it is well correlated with the particle rotational auto-correlation function. Detailed numerical analysis reveals that the physical origin of such unexpected directional motion is the imbalance of the frictional forces or hydrodynamic resistances experienced by the constituent atoms of the particle during the orientation regulation, which leads to a non-zero effective driving force along the initial orientation of the particle. This driving force only exists for a finite time interval before the onset of the diffusive Brownian motion. Therefore, it cannot be rationalized using the conventional theory of diffusive Brownian motion. Instead we derive theoretical formulae based on the observed microscopic picture that can well describe the mean displacement and the variance of this spontaneous directional motion. A virtual experiment is designed to demonstrate how such process could be detected and potentially applied in nanoscience. It is noted that the directional motion of shaped particles cannot result in directional drifting flow in equilibrium systems, because the ensemble-averaged effect leads to zero drifting flow at any time due to the equal probability distribution of the particle orientations. Our finding clearly shows that the directional motion process occurs in nanoscale space and within pico- to nanoseconds. It is completely different from the isotropic diffusive Brownian motion generally studied at much larger time and length scales. This study is expected to make important contribution to developing a complete microscopic theory for the motions of variously-shaped particles that can cover all interested time scales from ballistic to diffusive regime.

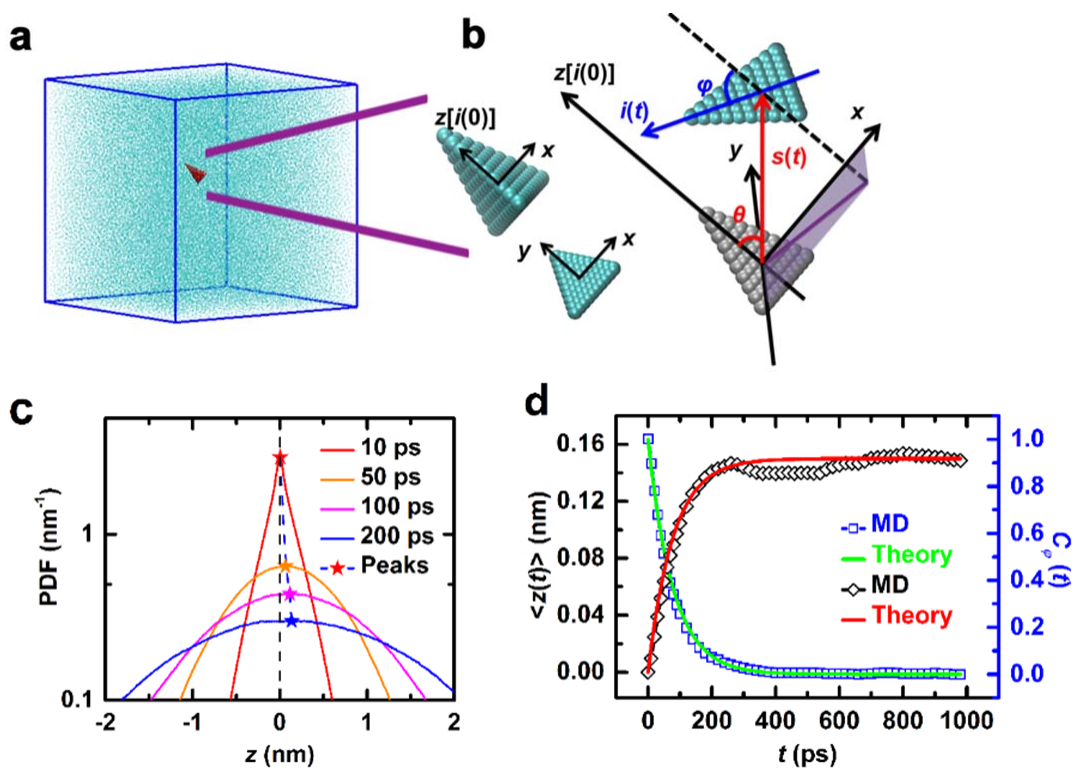

**Figure 1** (a) Snapshot of the simulation system consisting of a model nanoparticle shaped as a triangular pyramid in dilute solution. The cyan points represent the solvent particles and the shaped nanoparticle is red highlighted. (b) A 3D Cartesian coordinate frame defined for a target model nanoparticle with the origin point fixed at the initial location of its center of mass and the $z$-axis aligned along its initial orientation $\mathbf{i}(0)$ at time $t = 0$. The other two axes are set on the triangular base of the nanoparticle. The orientation of the shaped nanoparticle at time $t$, $\mathbf{i}(t)$, is defined by the unit vector (blue) pointing from the center of the triangular bottom to the top of the pyramid. The displacement of the shaped nanoparticle at time $t$ is the vector $\mathbf{s}(t)$. The position angle $\theta$ is between $\mathbf{s}(t)$ and the $z$-axis $[\mathbf{i}(0)]$, and the orientation angle $\varphi$ is between $\mathbf{i}(t)$ and the $z$-axis $[\mathbf{i}(0)]$. (c) Probability density function (PDF) of the center of mass (CoM) position projected onto the $z$-axis at different time $t$, where the stars mark the peaks of the PDFs. (d) $z$-component of the mean position of the CoM with respect to time (open black diamonds). The red line is the theoretical curve based on Eq. (6) with $A = 0.15$ nm. The open blue squares are the auto-correlation function of the particle orientation $C_\varphi(t)$ obtained from MD simulation. The green curve represents an exponential relaxation function based on Eq.(S2.2) in SI (Sec. 2) with the rotational diffusion coefficient $D_r = 6.48 \times 10^{-3}$ rad$^2$ ps$^{-1}$.

**Results from MD simulations**

Our simulation system consists of a single model nanoparticle shaped as triangular pyramid dispersed in a solvent of small Lennard-Jones particles with periodic boundary conditions applied in all three directions, as shown in Fig. 1(a) and described in the Simulation Method section. The MD simulation algorithms we employed have been widely used in the study of dynamics of molecules at nanoscale.[33–36] As detailed in SI (Sec. 9), the size of the simulation box is large enough and then the finite size effect[37] is negligible. The 3D Cartesian coordinate frame used for the particle motion analysis is displayed in Fig. 1(b) where the origin point and z-axis are built based on the initial location of the CoM and orientation **i**(0) of the shaped nanoparticle at a given reference time $t = 0$. Just like the calculation of any time correlation functions, each MD simulation time step can be taken as a time origin. All simulation results presented in this work are ensemble-averaged values obtained by using time origins separated by 1 ps along the MD trajectory after the MD system has reached equilibrium state, see Simulation Method. Figures 1(c) shows the probability density function (PDF) of the CoM position projected on the z-axis for the shaped nanoparticle. We note that the peak position of the PDF shifts toward the positive z-direction as time increases, indicating a directional motion.

We calculate the mean position or displacement of the CoM of the model particle on the z-axis [Fig. 1(d)] as a function of time,

$$\langle z(t) \rangle = \langle \mathbf{s}(t) \cdot \mathbf{i}(0) \rangle = \langle |\mathbf{s}(t)| \cos\theta \rangle, \qquad (1)$$

where $\theta$ is the angle between the displacement vector **s**(t) and the initial orientation (z-axis) of the nanoparticle as defined in Fig.1(b). As shown in Fig. 1(d), <z(t)> increases rapidly when $t < 100$ ps and gradually reaches a plateau value (0.15 nm) after 200 ps. This behavior is essentially different from the conventional theory of diffusive Brownian motion where the mean displacements of particles under no external influence are supposed to be zero, as already observed for spherical or ellipsoidal particles[19]. We have also computed the auto-correlation function of the particle orientation, $C_\varphi(t)$, which characterizes the particle rotational relaxation and is defined as

$$C_\varphi(t) = \langle \mathbf{i}(t) \cdot \mathbf{i}(0) \rangle = \langle \cos[\varphi(t)] \rangle \approx e^{-2D_r t}, \qquad (2)$$

where $\varphi(t)$ is the angle between the unit vectors **i**(t) and **i**(0) representing the particle orientations at time t and time zero. Conventionally, $C_\varphi(t)$ is supposed to decay exponentially and the rotational diffusion coefficient $D_r$ (= $6.48 \times 10^{-3}$ rad$^2$ ps$^{-1}$) can be obtained by linearly fitting the mean square angle of the particle orientation, $<\varphi^2(t)>$, to the Einstein relation[18] (see Sec. 2 in SI). As shown in Fig. 1(d), apart from the very first dozens of picoseconds where the inertial effect dominates (more details in Sec. 2 of SI), the theoretical prediction of $C_\varphi(t)$ [eq.(2)] agrees with the MD data very well up to the timescale of 1 ns. Comparing the $C_\varphi(t)$ and <z(t)> data in Fig.1(d) reveals an intrinsic correlation between the rotational and translational motion, as discussed below.

**Physical origin of the directional motion**
In order to understand the underlying physics of the spontaneous directional motion of the model particle, we compute the ensemble-averaged force <$F_z(t)$> acting on the nanoparticle along the z-axis by surrounding solvent particles. The result is shown in Fig. 2(a). It is clear that <$F_z(t)$> does have a positive value at early times. More specifically this force first increases from zero, reaches its maximal value at ~2 ps and then decreases. It becomes negative at 12 ps, reaches the minimal value at 17 ps and then gradually approaches zero. As a direct consequence, the z-components of both the mean translational velocity, <$v_z(t)$>, [Fig.2(b)] and the mean displacement, <z(t)>, [Fig.1(d)] possess positive values for over 200 ps.

Now we focus on the physical origin of the driving force for the spontaneous directional motion by analyzing the ensemble-averaged force <$F_z^i(t)$> acting on every individual constituent atom of the shaped nanoparticle, where $i$ is the serial number of the atom under investigation. Apparently $<F_z^i(t)>$ only results from the fluctuating forces owing to the collision with the surrounding solvent molecules. Thus, $<F_z^i(t)>$ depends on the movement of each atom and can be written as a function of both its transitional and rotational velocities (detail in Sec. 3 of SI). Then solving the Newton's second law, which is a differential equation of velocity, we can obtain the mean velocity $<v_z(t)>$,

$$\langle v_z(t) \rangle = C_2 R(t) e^{-C_1 t} \text{ with } R(t) = \int_0^t Q(\xi) e^{C_1 \xi} d\xi, \qquad (3)$$

where

$$C_1 = \frac{\sum_{i=1}^N \lambda^i}{m}, \quad C_2 = \frac{\sum_{i=1}^N \lambda^i r_o^i}{m} \text{ and } Q(t) = -\frac{d}{dt} C_\varphi(t). \qquad (4)$$

In Eq. (4), $m = Nm_{LJ}$ is the mass of the nanoparticle and $m_{LJ}$ is the mass of each constituent atom defined in the Simulation Method section. $\lambda^i$ is the effective frictional coefficient of the atom located at different sites of the nanoparticle structure and $r_o^i$ is the projection of the atom position vector $\mathbf{r}^i$, (from the particle CoM to the center of the $i$th atom) on the particle orientation axis and can be obtained directly from the construction of the particle structure. Actually a more complicated empirical expression for $C_\varphi(t)$ (see Sec. 2 in SI), rather than Eq. (2), has been used in Eq. (4) for getting the best agreement between the theoretical calculated $<v_z(t)>$ and later the mean force $\langle F_z(t) \rangle$ with the simulation data.

The mean force acting on the particle is thus

$$\langle F_z(t) \rangle = mC_2 [Q(t) - C_1 R(t) e^{-C_1 t}]. \qquad (5)$$

This formula further clarifies that the mean force on the nanoparticle can be nonzero only when $C_2 \neq 0$ and $[Q(t) - C_1 R(t)\exp(-C_1 t)] \neq 0$. The contribution of the asymmetrical architecture of the particle is carried by $C_2$, since it possesses the position vector for every atom. The terms in the square bracket are all rotational motion related. It indicates that the key reason for the non-zero net force $<F_z(t)>$ lies in the imbalance of the effective frictional forces or hydrodynamic resistances ($C_2 \neq 0$) acting on the constituent atoms during the particle orientation regulation. This can only happen for particles with asymmetric structures. If the geometric shape of the particle is symmetric, the effective frictional forces acting on all atoms are well balanced ($C_2 = 0$) and no directional motion can be observed, see examples given in SI (Sec. 7). On the other hand, the rotational motion also plays the important role. If we fix the orientation of the particle [$C_\varphi(t) = 1$], $Q(t)$ and $R(t)$ in Eq. (5) both equal to zero. Then the mean force even for an asymmetrically shaped nanoparticle with $C_2 \neq 0$ is zero either. Therefore, the key of nonzero mean force is asymmetry, while the rotation is also required.

It is rather tedious to compute $C_1$ and $C_2$ from estimating the effective frictional coefficients $\lambda^i$ of all the atoms. For practical purpose, we estimate $C_1$ and $C_2$ from the best fitting of the simulation data on $<F_z(t)>$ and $<v_z(t)>$ to Eqs. (3) and (5). As shown in Fig. 2, these theoretical formulae provide good description of the simulation results. In SI (Sec. 6) we also use a smaller model nanoparticle consisting of 10 atoms to demonstrate that the fitted $C_1$ and $C_2$ values are in very good agreement with those directly calculated using Eq. (4).

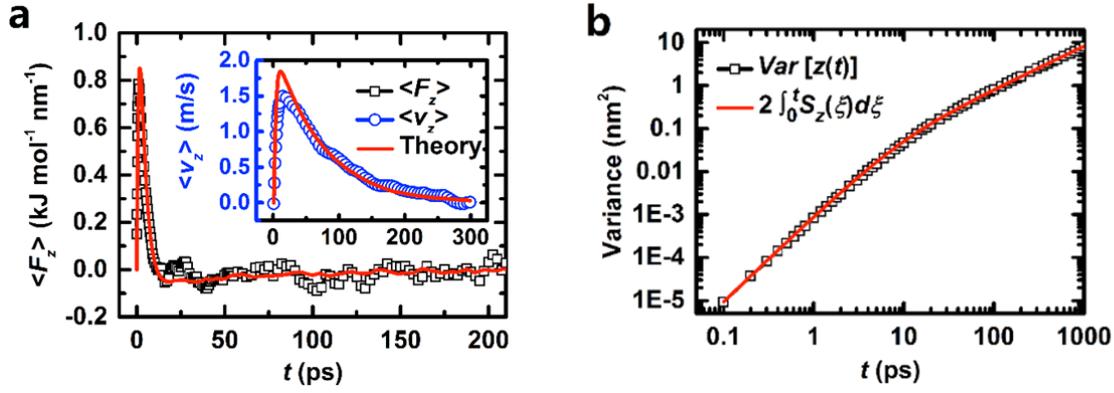

**Figure 2** (a) Mean force experienced by the shaped nanoparticle studied in Fig. 1 and its mean velocity (inlet), both along the *z*-axis. The solid curves are the best fitting of the simulation results (symbols) to the theoretical formulae in Eqs. (5) and (3), respectively. (b) Variance of the displacement of the particle along the *z*-axis, $Var[z(t)]$, as a function of time. The solid red curve is the numerical fitting using Eq. (8).

**Theoretical description of the directional motion**

The mean position $<z(t)>$ of the shaped nanoparticle can be obtained by integrating the mean velocity [Eq. (3)],

$$\langle z(t) \rangle = \int_0^t \langle v_z(\xi) \rangle d\xi = \frac{C_2}{C_1}\left[1 - C_\varphi(t) - R(t)e^{-C_1 t}\right]. \tag{6}$$

Equation (6) again suggests that the directional motion of the particle is determined by $C_2$, which associates with the driving force. Since the term $R(t)\exp(-C_1 t)$ has a peak height of only about 0.02 s$^{-1}$ at $t \approx 10$ ps and then decays rapidly to zero, we can simply write Eq. (6) at larger time scales ($t > 10$ ps) as

$$\langle z(t) \rangle \approx A\left[1 - C_\varphi(t)\right] \text{ with } A = \lim_{t \to \infty} \langle z(t) \rangle = \frac{C_2}{C_1}. \tag{7}$$

Equation (7) clearly elucidates the simple correlation between rotational motion [$C_\varphi(t)$] and the transitional motion [$<z(t)>$]. It is shown in Fig. 1(d) that this expression provides very good description of the simulation data on $<z(t)>$. We note that Charkrabarty et al. obtained a similar relationship between the mean displacement of the tracking point (CoB) and the particle rotational autocorrelation function, e.g., see Eqs. (15a, b) in Ref. [21]. But their displacement did not stand for the coupled motion of mass and the correlation function they used takes a single exponential form, since their calculations were carried out in the diffusive Brownian motion regime.

Since the mean displacement of the shaped nanoparticle (first moment of the PDF of its position) along the *z*-axis is non-zero due to directional motion, i. e., $<z(t)> \neq 0$, we choose to study the second moment of the PDF, namely the variance $Var[z(t)]$ of the particle displacement, rather than the mean square displacement (MSD) which conventionally assumes $<z(t)> = 0$. Based on the observed translation-rotation coupling picture, this quantity can be described by an empirical expression,

$$Var[z(t)] \equiv \langle [z(t) - \langle z(t) \rangle]^2 \rangle = 2\int_0^t S_z(\xi)d\xi, \tag{8}$$

where

$$S_z(t) = D_z\left(1 - e^{-t/\tau_z}\right) \text{ with } \tau_z = \frac{mD_z}{k_B T}. \tag{9}$$

As shown in Fig. 2(b), Equation (8) fits the data from MD simulation very well with the one-dimensional (1D) diffusion coefficient $D_z = 4.12 \times 10^{-3}$ nm$^2$ ps$^{-1}$ and the velocity relaxation time $\tau_z = 4.36$ ps. Considering that $<[z(t)-<z(t)>]^2> = <[z(t)]^2> - [<z(t)>]^2$ and $<[z(t)]^2>$ is the 1D MSD, equation (8) can be justified in the two time limits. At very small time scales ($t < \tau_z$), equation (8) is approximated by $S_z(t) \approx D_z t/\tau_z$ and so $Var[z(t)] \approx <v_T^2>t^2$ where $v_T$ is the velocity of thermal motion based on the equipartition theorem, $k_B T = m<v_T^2>$. It follows that the MSD of the particle $<[z(t)]^2> \approx <v_T^2>t^2 + [<z(t)>]^2$, reflecting the ballistic motion behavior plus a spontaneous directional motion. On the other hand, at large enough time scales, $Var[z(t)] \approx 2D_z t$ and the MSD $<[z(t)]^2> \approx 2D_z t + [<z(t)>]^2$. Owing to the limited saturation value of the mean displacement $<z(t)>$, the Einstein relationship, $<[z(t)]^2> \approx 2D_z t$, is recovered at large time. We note that the results in Fig. 2(b) are consistent with the 3D MSD of the shaped nanoparticle calculated in the laboratory coordinate system as shown in Fig. S2.2 of SI (Sec. 2).

**Potential detection and application of the directional motion**

The impact of this spontaneous directional motion can be detected at time and length scales much larger than the process itself. Figure 3 demonstrates one example of how the spontaneous directional motion of a shaped nanoparticle can affect the probabilities of this particle to meet targets located differently in relation to its initial orientation direction.

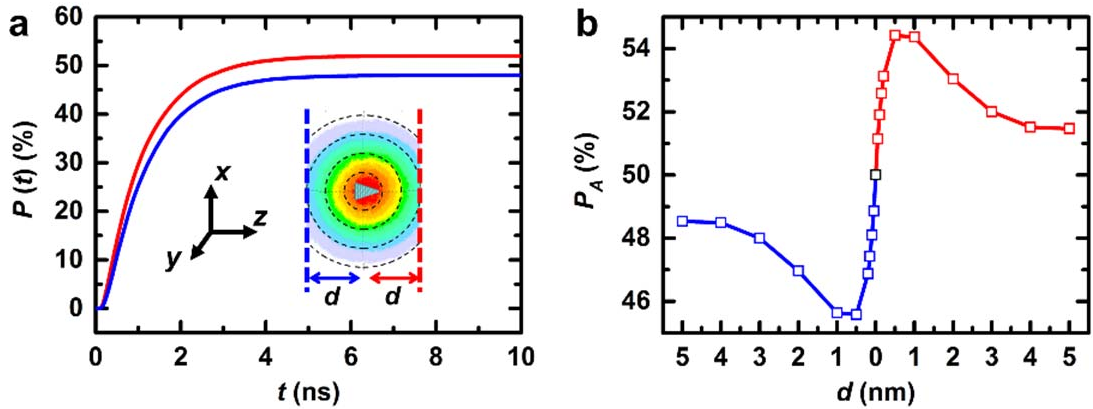

Figure 3. (a) Probabilities for a pyramid-shaped nanoparticle absorbed by one of the two boundaries with respect to time $t$ when the planar absorbing boundaries are located at $z = \pm d$ with $d = 3$ nm, respectively. The colored background of the inset shows the probability distribution function of the CoM position of the particle projected on the $x$-$z$ plane. (b) Saturated values of the absorption probability with respect to the initial distance $d$ from the CoM of the model particle to the boundaries.

As shown in Fig. 3, a pyramid-shaped particle is introduced into the solvent of small LJ particles with its center of mass located at the origin of the Cartesian coordinate frame and its principal orientation aligned along the positive $z$-direction. Two planar absorbing boundaries are placed in parallel to the $x$-$y$ plane and located at $z = \pm d$, respectively, see the inset in Fig. 3(a). The particle is released at time $t = 0$ to allow for free motion. When the CoM of the particle reaches any one of the two absorbing boundaries for the first time, it is considered to be absorbed. Fig. 3(a) clearly shows that the spontaneous directional motion of the shaped nanoparticle has led to a higher probability for it to be absorbed by the boundary located at $z = d$, *i. e.*, pointed to by its original orientation. The plateau values of the absorbed probabilities of the particles by the two boundaries differ by about 4% for $d = \pm 3$ nm. It is noted that for model particles of this size the maximum mean directional displacement is only $A \approx 0.15$ nm (see Fig. 1(d)), which is much smaller than the initial distance $d$ of the particle CoM to the boundaries. As shown in Fig. 3(b), the difference in the absorption probabilities reaches a peak value in between $d = 0.5$ to 1 nm, then gradually decreases with the increase of $d$, but still does not vanish after $d = 5$ nm ($> h = 1.68$nm $>> A$). As implied in Fig. S8.1 in SI (Sec. 8), the effect by the spontaneous directional motion can be amplified by using particles of larger sizes or of different asymmetric shapes. It is thus possible to detect this process in experiments by employing well-selected shaped particles/molecules with embedded magnetic or electric dipole moments and applying external field to achieve the initial particle orientation.

On the other hand, the diffusion-reaction problems are encountered in many biological, chemical and physical processes, such as gene delivery and viral infection which involve the kinetically limited search processes for nuclear sites by nucleic acids[38,39]. The intracellular signaling between macromolecules or organelle in cell is also carried out by small signal molecules approaching the receptors within several nanometers where the probability of meeting targets is the key[9,39,5]. Since all of the involved molecules are of asymmetric shapes, to determine the role of the spontaneous directional motion in the dynamics of these processes remains to be an open and challenging research subject. Another interesting topic would be the design of novel bio-sensors by taking advantage of the biased meeting probabilities of the asymmetric particles in relation to their initial orientation as shown in Fig.3.

**Conclusion**
We have shown by molecular dynamics simulations and theoretical analyses that shaped solute molecules/particles with broken central symmetry in dilute solutions undergo spontaneous directional motion processes along their original orientation directions. Although in conventional theories particle drifting can only happen in the presence of external situational factors, such as concentration or temperature gradient, our simulation results reveal that for asymmetrically shaped nanoparticles, the directional motion is spontaneous and so takes place even without any external interference. The physical origin of the effective driving force for the spontaneous directional motion can be attributed to the imbalanced fluctuating forces acting on the particles by surrounding solvent molecules during the particle rotational relaxation. The mean displacement of the particle gets saturated after a sufficiently long time at which the auto-correlation function of the particle orientation decays to zero. The impact of the spontaneous directional motion may be more significant for asymmetrically shaped particles of larger sizes (e.g., colloidal particles) in the sense of absolute magnitude, since the saturated value of the mean displacement grows with the size of the particles.

In this paper, we only consider the spontaneous directional motion along one direction (initial particle orientation), while the non-zero mean displacement can actually occur in multiple directions for particles or molecules with broken central geometric symmetry within a finite time. Moreover, other types of asymmetries, such as the inhomogeneous mass distribution, charge distribution, hydrophobic/hydrophilic domains, will also induce the directional motion, which will be discussed in later works.

A further remark we would like to make is that the observed spontaneous directional motion will not lead to a perpetual mobile that violates the second law of thermodynamics. As shown above, the directional motion of shaped nanoparticles only take place in the directions along their initial orientations. In equilibrium systems, the orientations of the nanoparticles have equal probability in all directions. After averaging over all possible initial orientations of the particles, the mean displacement is zero and so there is no directional flow in the system, which is consistent with the statistical mechanic principles. In the SI (Sec. 10), we have demonstrated that one cannot extract mechanical energy from a thermal bath by restraining the orientations of the particles. To make use of the spontaneous directional motion phenomenon, external energy input is required. However, for the case that we only focus on the motion of a single molecule within a finite time, just as the example of we presented above, biased moving probability and the spontaneous directional motion will definitely affect various physical, chemical and biological processes happened within spaces comparable to molecule/particle sizes and within finite timescales.

We note that extensive physical, chemical and even biological processes, including protein conformation changes[4,40], occur in a finite time, e.g., nano-, pico- and even femto-seconds. Therefore, our finding may contribute to understanding the microscopic mechanisms of various kinetic processes and their practical applications, such as chemical separation[28,41], sensing[42–44] and drug delivery[45,46], especially at short timescales. Our finding clearly shows that the motions of asymmetrically shaped molecules/particles in the nanoscale space and within pico- to nanoseconds are completely different from the conventional isotropic diffusive Brownian motion picture. We hope that this study can inspire the development of a complete theoretical framework that can describe the motions of variously-shaped particles in solutions over a whole range of timescales from ballistic to diffusive regime.

**Simulation Method**
The model nanoparticle we simulated was shaped as triangular pyramids with height $h = 1.68$ nm, as shown in Fig. 1(a), where three side surfaces were identical isosceles triangles with the angle of 36° and the bottom surfaces were regular triangles. The nanoparticle was built by bonding 220 Leonard-Jones (LJ) particles. We put the model nanoparticle in a cubic box of dimensions 18 nm × 18 nm × 18 nm with periodic boundary conditions filled with 97 306 LJ particles as solvent. In SI (Sec. 9) we demonstrated that the finite size effects are negligible for the simulation results on the directional motion of the model particles. All LJ particles had the same mass of $m_{LJ}$=12.011 u and the same force field parameters ($\sigma = 0.375$ nm, $\varepsilon = 0.439$ kJ mol$^{-1}$). The cut-off distance for van de Waals (vdW) interactions was set to 1.3 nm. There was no Coulomb interaction in this system. The temperature was maintained at 300 K by velocity-rescale thermostat[47]. A time step of 2 fs was used, and the neighbor list was updated every 10 steps. Using Gromacs 4.6[48] software, we performed 5 independent simulation runs for each system containing a single model particle immersed in solvents, starting from different initial configurations. Each system was first equilibrated for 20 ns and then ran for another 200 ns for analysis. The calculation of the PDFs of the CoMs and consequently the mean displacements of the model particles were carried out by taking time origins separated by 1ps along the MD trajectory. Thus we had about 1 million samples of each model nanoparticle for statistical analysis.

\* Corresponding author. Email: fanghaiping@sinap.ac.cn
† Corresponding author. Email: zuowei.wang@reading.ac.uk

**Acknowledgements**
We thank Prof. JI Qing and Prof. HU Jun for helpful discussions. Z. Wang also acknowledge Alexei Likhtman, Alex Lukyanov and Eugene Terentjev for valuable discussions. This work was supported by the National Natural Science Foundation of China (Grant Nos. 10825520, 11422542, 11175230 and 11290164), the Key Research Program of Chinese Academy of Sciences (Grant No. KJZD-EW-M03), the Deepcomp7000 and ScGrid of Supercomputing Center, the Computer Network Information Center of Chinese Academy of Sciences and the Shanghai Supercomputer Center of China.

**Author Contributions**
H.F., Z.W. and N.S. contributed the idea and designed the project. N.S. performed the numerical simulations. H.F., Z.W. and N.S. carried out most of the theoretical analysis, and wrote the paper. Y.T., P.G. and R.W. performed some theoretical analysis. All authors discussed the results and commented on the manuscript.

**Additional information**
**Supplementary Information** is available in the online version of the paper.

**Competing Financial Interests:** The authors declare that they have no competing financial interests.